\documentstyle[12pt]{article}
\begin{document}
\baselineskip=22.pt plus 0.2pt minus 0.2pt
\lineskip=22.pt plus 0.2pt minus 0.2pt

\setcounter{equation}{0}

\centerline{\Large\bf A New Dimensionally Reduced Effective Action}
\centerline{\Large\bf for QCD at High Temperature}

\vskip 0.8cm

\centerline{\large Scott Chapman}

\centerline{\em Institut f\"ur Theoretische Physik, Universit\"at
Regensburg,}
\centerline{\em D-93040 Regensburg, Germany}

\vskip 0.8cm
\noindent\underbar{\bf Abstract:}
New terms are derived for the three-dimensional effective action of
the static modes of pure gauge SU(N) at high temperature.  In previous
works, effective vertices have been obtained by evaluating diagrams
involving 2, 3 or 4 external static gluons with one internal nonstatic
loop.  I take a somewhat different approach by making a covariant
derivative expansion of the one loop effective action for the static
modes, keeping all terms involving up to six covariant derivatives.
The resulting effective action is manifestly invariant under spatially
dependent gauge transformations and contains new 5- and 6-point
effective vertices.

\vskip .1 cm

\vspace*{.5in}

\section{Introduction}

Several years ago it was realized that for sufficiently high
temperatures, the dominant infrared behavior of four-dimensional QCD
could be described by a static three-dimensional theory
\cite{app1}. The basic idea is the following: In the imaginary time
formalism of finite temperature gauge field theory, the gauge (and
ghost) fields must be periodic in imaginary time $\tau$ with period
$1/T$ \cite{kap}. These fields ($A^a_\mu(x)$) can thus be expanded
into fourier modes $A^a_{\mu,n}({\bf x})\exp(i2n\pi T\tau)$, each of
which has a propagator of the form $[{\bf p}^2 + (2n\pi T)^2]^{-1}\;$.
For any nonzero $n$, $2n\pi T$ acts as a mass which causes the $n$th
mode to become infinitely ``heavy'' as the temperature goes to
infinity.  Since infinitely heavy particles decouple from zero
temperature field theories \cite{app2}, it has been argued that the
same should happen to nonstatic $n\ne 0$ modes in high temperature
field theories \cite{app1}.  This would imply that at sufficiently
high temperatures, four dimensional pure gauge SU(N) would reduce to
three dimensional SU(N) coupled to an adjoint scalar $(A_0^a)$:
\begin{equation}
\int d^4x(F_{\mu\nu}^a(x))^2 \rightarrow (1/T)\int d^3x[
(F_{ij}^a({\bf x}))^2 + 2(D_i^{ab}A_0^b({\bf x}))^2]
\label{i1}
\end{equation}

Unlike zero temperature field theories with heavy particles, however,
this kind of complete decoupling does not in fact occur for QCD even
at infinite temperature \cite{land}.  Rather, the nonstatic modes make
themselves felt within loops by generating an infinite number of
effective vertices for the static modes which cannot in general be
ignored.  Consequently, many authors have extended the above
3-dimensional theory by including the leading corrections to 2-, 3-
and 4-point static vertices which come from diagrams involving one
internal nonstatic loop \cite{land,nadk,rei}.  Encouragingly, lattice
results indicate that 3-dimensional QCD extended in this way
reproduces the full 4-dimensional QCD remarkably well for temperatures
above 2.5 times that of the phase transition \cite{kark}.  For lower
temperatures, however, the nonstatic modes contribute more
significantly, thus causing the simple actions proposed so far to
become inadequate to describe the full theory.  It would be
interesting to see if better agreement could be generated at lower
temperatures by including more of the induced static vertices in the
effective theory.  I take a first step towards answering that question
by deriving an effective action which contains new 5- and 6-point
static vertices, as well as derivative corrections to known 2-, 3- and
4-point vertices.  These derivative corrections essentially constitute
new terms in the $(|{\bf p}|/T)$ infrared expansion proposed in
\cite{rei}, where ${\bf p}$ represents the external momentum of a
given effective vertex.

I begin this work by making a covariant derivative expansion of the
one loop effective action of SU(N) in the presence of a general
background field, keeping all terms with up to six covariant
derivatives.  I then specialize this general result (eqn.
\ref{7}) to the problem at hand by considering only static background
fields and restricting the one loop integration to nonstatic
fluctuations.  The resulting 3-dimensional effective action is
manifestly invariant under any spatially-dependent gauge transformation.

\section{Covariant Derivative Expansion}

The one loop contribution to effective action of pure gauge SU(N) can
be calculated by evaluating the following functional determinants
\cite{chap,dyak}:
\begin{equation}
S_{\rm eff}^{(1)} = \ln[\frac{\det(-D^2)}{\det(-\partial^2)}]
-{\textstyle\frac{1}{2}\,}\ln[\frac{\det(-D^2\delta_{\mu\nu}-2[D_\mu,D_\nu])}
{\det(-\partial^2\delta_{\mu\nu})}]\;,
\label{1}
\end{equation}
where $D_\mu^{ab}=\partial_\mu\delta^{ab}-gf^{abc}A_\mu^c({\bf x})$.
To evaluate these determinants, we use the identity:
\begin{equation}
\ln[\frac{\det(K)}{\det(K_0)}] = -\int_0^\infty \frac{dt}{t}
[{\rm Tr}(e^{-tK}-e^{-tK_0})]
\label{1a}
\end{equation}
for operators $K$ and $K_0$.

Using plane waves to take the functional trace of the ghost operator,
we arrive at the following expression \cite{chap}:
\begin{equation}
\ln\det(\frac{-D^2}{-\partial^2}) = -\int_0^\infty \frac{dt}{t}
\int \frac{d^4xd^3p}{(2\pi)^3}T\sum_n e^{-p^2t}
{\rm tr}\{\exp[(D^2+2iD_\alpha p_\alpha) t]{\bf 1}-1\}\;,
\label{4}
\end{equation}
where the remaining trace tr is taken only over the color indices in
the adjoint representation.  At finite temperature, the $x_0$ integral
in $d^4x$ is from 0 to $1/T$, and instead of integrating over $p_0$,
one must sum over the modes $p_0=2n\pi T$.  A {\bf 1} has been
included at the end of the equation to emphasize the fact that the
exponential operator acts on unity; so that, for example, any term in
the expansion of the exponent with a $\partial_\alpha$ all the way to
the right will vanish.

With the help of the appendix, we expand the exponent, keeping all
terms involving only $D_0$ or up to six covariant derivatives of any
kind.  After integrating over $d^3p$, we obtain:
\begin{eqnarray}
&\,&\ln\det(\frac{-D^2}{-\partial^2}) = \int d^4x V_{\rm eff}
-\int_0^\infty \frac{dt}{t}
\int \frac{d^4x}{(4\pi t)^{3/2}}(\frac{T}{180})\sum_n e^{-p_0^2t}
\nonumber \\
&\,&\;\;\;\times\,{\rm tr}\,\{\,
15t^2[D_\mu,D_\nu]^2
+30t^2(1-2p_0^2t)([D_i,D_0]^2 +\{[D_i,[D_i,D_0]],D_0\})
\nonumber \\
&\,&\;\;\;\;\;
+t^3(2{\cal O}_1-3{\cal O}_2)
+3t^3(1-2p_0^2t)[2{\cal P}_1 -2({\cal P}_4+{\cal P}_{13})
-{\cal P}_6]
\nonumber \\
&\,&\;\;\;\;\;+3t^3(1-2p_0^2t)[-4({\cal Q}_1+{\cal Q}_9)
-2({\cal Q}_3+2{\cal Q}_9) +5({\cal Q}_7+2{\cal Q}_{12})
+({\cal Q}_8+{\cal Q}_{14})]
\nonumber \\
&\,&\;\;\;\;\;+2t^3(3-12p_0^2t+4p_0^4t^2)
[-{\cal P}_{13} -6{\cal Q}_9 +15{\cal Q}_{12} +5{\cal Q}_{13}
+2{\cal Q}_{14}]\,\}\,{\bf 1}\;,
\label{5}
\end{eqnarray}
where the sixth order ${\cal O}$, ${\cal P}$, and ${\cal Q}$ operators
are defined in the appendix, and the effective potential is given by:
\begin{equation}
V_{\rm eff} = -\int_0^\infty \frac{dt}{t}
\frac{T}{(4\pi t)^{3/2}}\sum_n
{\rm tr}\{\exp[(D_0+ip_0)^2t]-e^{-p_0^2t}\}{\bf 1}\;.
\label{5a}
\end{equation}

The gauge field determinant can be calculated using techniques similar
to those used for the ghost determinant.  One obtains:
\begin{eqnarray}
&\,&\ln[\frac{\det(-D^2\delta_{\mu\nu}-2[D_\mu,D_\nu])}
{\det(-\partial^2\delta_{\mu\nu})}]
=4\ln\det(\frac{-D^2}{-\partial^2})
\nonumber \\
&\,&\;\;\;-\int_0^\infty \frac{dt}{t}
\int \frac{d^4x}{(4\pi t)^{3/2}}\frac{T}{3}\sum_n e^{-p_0^2t}
\,{\rm tr}\{\,-6t^2[D_\mu,D_\nu]^2 +2t^3{\cal O}_2
\nonumber \\
&\,&\;\;\;+t^3(1-2p_0^2t)[-4{\cal P}_1
+ 2({\cal P}_4+{\cal P}_{13}) +4({\cal Q}_3+2{\cal Q}_9)
-6({\cal Q}_7+2{\cal Q}_{12})] \, \}\,{\bf 1}\;.
\label{6}
\end{eqnarray}
Note that the factor of 4 in the first line arises simply from the
trace over $\delta_{\mu\nu}$.

For convenience, we now switch to matrix notation, defining
$(I_c)^{ab}=-if^{abc}$, $A_0=I_cA_0^c$, $F_{\mu\nu}=I_cF_{\mu\nu}^c$,
$(D_\mu F_{\nu\rho})=I_cD_\mu^{cd}F_{\nu\rho}^d$, etc.  The one loop
contribution to the effective action then takes the following form:
\begin{eqnarray}
S_{\rm eff}^{(1)} &=& -\int d^4x V_{\rm eff} \;+\;
g^2\int\frac{d^4x}{(4\pi)^{3/2}}T\sum_n \int_0^\infty
\frac{dt}{\sqrt{t}}e^{-p_0^2t}\,{\rm tr}\,\{\,
{\textstyle\frac{11}{12}\,}F_{\mu\nu}^2
\nonumber \\
&-&{\textstyle\frac{1}{6}\,}
(1-2p_0^2t)[F_{\mu 0}^2 +2A_0(D_\mu F_{\mu 0})]
+{\textstyle\frac{1}{90}\,}tig
F_{\mu\nu}F_{\nu\rho}F_{\rho\mu}
-{\textstyle\frac{19}{60}\,}t
(D_\mu F_{\mu\nu})^2
\nonumber \\
&+&t(1-2p_0^2t)[-{\textstyle\frac{19}{30}\,}ig
F_{0\mu}F_{\mu\nu}F_{\nu 0}
-{\textstyle\frac{3}{10}\,}(D_0F_{0\nu})(D_\mu F_{\mu\nu})
+{\textstyle\frac{1}{60}\,}(D_\mu F_{\mu 0})^2
\nonumber \\
&\,&\;\;\;\;\;\;\;\;\;\;-{\textstyle\frac{1}{15}\,}
igA_0(D_\mu F_{\mu\nu})F_{0\nu}
+{\textstyle\frac{19}{30}\,}igA_0(D_0F_{\mu\nu})F_{\mu\nu}
\nonumber \\
&\,&\;\;\;\;\;\;\;\;\;\;\;
-{\textstyle\frac{11}{12}\,}g^2A_0^2F_{\mu\nu}^2
-{\textstyle\frac{1}{30}\,}A_0(D_\mu^2D_\nu F_{\nu 0})]
\nonumber \\
&+&t(3-12p_0^2t+4p_0^4t^2)[{\textstyle\frac{1}{90}\,}
(D_0F_{0\mu})^2
-{\textstyle\frac{1}{15}\,}igA_0(D_0F_{0\mu})F_{0\mu}
\nonumber \\
&\,&\;\;\;\;\;\;\;\;\;\;
+{\textstyle\frac{1}{6}\,}g^2A_0^2F_{\mu 0}^2
+{\textstyle\frac{1}{9}\,}g^2A_0^3(D_\mu F_{\mu 0})
-{\textstyle\frac{2}{45}\,}A_0(D_0^2D_\mu F_{\mu 0})]\}
\label{7}
\end{eqnarray}
Only the terms at the end of the first and second lines of the above
equation are relativistically covariant.  Not surprisingly, these are
the only terms which remain at zero temperature \cite{dyak}, as can
easily be seen if one replaces the sum over $n$ by an integral over
$p_0$.  On the other hand, all of the terms in (\ref{7}) are invariant
under purely spatially-dependent gauge transformations and rotations.

\section{Effective Action for Static Modes}

Just as at zero temperature, the coefficient of $F_{\mu\nu}^2$ is
ultraviolet divergent and needs to be regulated.  This can easily be
handled by zeta function regularization \cite{dyak,hawk}:
\begin{eqnarray}
[T\sum_{n\ne 0}\int\frac{dt}{\sqrt{t}}e^{-p_0^2t}]_{\rm reg}
&=& T\sum_{n\ne 0}\frac{d}{d\epsilon}
\{\frac{\Lambda^{2\epsilon}}{\Gamma(\epsilon)}\int
dt\,t^{\epsilon-1/2}e^{-p_0^2t}\}_{\epsilon\rightarrow 0}
\nonumber \\
&=&\frac{1}{\pi}\frac{d}{d\epsilon}\{(\frac{\Lambda}{2\pi T})^{2\epsilon}
\frac{\Gamma(\epsilon+{\textstyle\frac{1}{2}\,})}
{\Gamma(\epsilon)}\zeta(1+2\epsilon)\}_{\epsilon\rightarrow 0}
\nonumber \\
&=& \frac{1}{\sqrt{\pi}}\{\ln(\frac{\Lambda}{4T})+\gamma_E\}
\label{11}
\end{eqnarray}
where $\gamma_E\sim .577$ is Euler's constant, $\Lambda$ is the
ultraviolet regulator mass, and we have used the fact that we are
only summing over nonstatic $(n\ne 0)$ fluctuations.

The use of a regulator affects not only the divergent $F_{\mu\nu}^2$
term but also the first term on the second line of (\ref{7}), even
though that term is UV-finite (and would vanish without the regulator).
We note that for static fields:
\begin{equation}
{\rm tr}\int d^3x\,A_0(D_\mu F_{\mu 0})
= -{\rm tr}\int d^3x\,(D_i A_0)^2
= -{\rm tr}\int d^3x\,F_{\mu 0}^2
\label{7a}
\end{equation}
Using the regulator to perform the frequency sum and $t$ integration, we find:
\begin{equation}
T\sum_{n\ne 0}\frac{d}{d\epsilon}
\{\frac{\Lambda^{2\epsilon}}{\Gamma(\epsilon)}\int
dt\,t^{\epsilon-1/2}e^{-p_0^2t}(1-2p_0^2t)\}_{\epsilon\rightarrow 0}
= -\frac{1}{\sqrt{\pi}}
\label{7b}
\end{equation}

It is now possible to renormalize in the usual way by defining
renormalized quantities in terms of bare (B) quantities via
multiplicative constants:
\begin{equation}
A_0^a = Z_E^{-1/2}A_{0,B}^a\;,\;\;\;\;\;\;\;\;
A_i^a = Z_M^{-1/2}A_{i,B}^a\;,\;\;\;\;\;\;\;\;
g = Z_g^{1/2}g_B
\label{7c}
\end{equation}
Dimensional reduction works best if the Z's are chosen such
that all of the contributions to $F_{ij}^2$ and
$F_{i0}^2$ which arise from nonstatic loops are cancelled by
counterterms \cite{land}.  In our case, this can be achieved by choosing:
\begin{equation}
Z_g = Z_M = Z_E - \frac{g^2N}{12\pi^2} = 1 +
\frac{11g^2N}{24\pi^2}[\ln(\frac{\Lambda}{4T}) + \gamma_E]
\label{8}
\end{equation}
The running coupling g(T) is now uniquely determined and takes the
usual form (see for example \cite{land}).  I would also like to note
that a finite difference between the electric (E) and magnetic (M)
wave-function renormalization constants has similarly been found by
other authors \cite{land,nadk}.

For the rest of the terms in eqn. (\ref{7}), the same result is found
whether we sum over $n$ and integrate over $t$ before taking the
$\epsilon\rightarrow 0$ limit of the regulator, or if we do things the
other way around.  To simplify the computation, we therefore choose to
take the $\epsilon \rightarrow 0$ limit first, thus effectively
removing the regulator.  The resulting sums and integrations are:
\begin{equation}
\sum_{n\ne 0}\int dt \sqrt{t}\,e^{-p_0^2t}
=-{\textstyle\frac{1}{2}\,}\sum_{n\ne 0}
\int dt \sqrt{t}\,(1-2p_0^2t)e^{-p_0^2t}
= \frac{\sqrt{\pi}}{(2\pi T)^3}\zeta(3)
\label{9}
\end{equation}
\begin{equation}
\sum_{n\neq 0}\int_0^\infty dt\sqrt{t}\,(3-12p_0^2t+4p_0^4t^2)
e^{-p_0^2t} = 0
\label{10}
\end{equation}
where $\zeta(x)$ is the Riemann zeta function with $\zeta(3)\sim
1.202$.  Furthermore, for static background fields some of the terms
can be combined using:
\begin{eqnarray}
{\rm tr}\int d^3x \{(D_0F_{0\nu})(D_\mu F_{\mu\nu})\} &=&
2ig\,{\rm tr}\int d^3x \{A_0(D_\mu F_{\mu\nu})F_{0\nu}\}
\nonumber \\
{\rm tr}\int d^3x \{A_0(D_\mu^2D_\nu F_{\nu 0})\} &=&
{\rm tr}\int d^3x \{(D_\mu F_{\mu 0})^2\}
\nonumber \\
{\rm tr}\int d^3x \{A_0(D_0F_{\mu\nu})F_{\mu\nu}\} &=&
2\,{\rm tr}\int d^3x  \{A_0(D_\mu F_{\mu\nu})F_{0\nu} +
F_{0\mu}F_{\mu\nu}F_{\nu 0}\} \;.
\label{10a}
\end{eqnarray}

We now turn to the effective potential.  Because we are only
interested in static background configurations, the time derivatives
vanish from eqn. (\ref{5a}), and we are left with:
\begin{equation}
V^\prime_{\rm eff} = -\int_0^\infty \frac{dt}{t}
\frac{1}{(4\pi t)^{3/2}}T\sum_{n\ne 0}
{\rm tr}\,\{\exp[-(p_0 - gA_0)^2t]-e^{-p_0^2t}\}\;,
\label{12}
\end{equation}
where the prime on $V^\prime_{\rm eff}$ indicates that we are only
integrating over nonstatic fluctuations.  Since there are no spatial
derivatives in $V^\prime_{\rm eff}$ (or $V_{\rm eff}$),
spatially-dependent unitary matrices can be used within the trace to
transform $A_0({\bf x})$ into a real, diagonal matrix at each point in
space.  In Appendix D of
\cite{gpy}, the SU(N) effective potential for a diagonal background
field is calculated to be:
\begin{eqnarray}
V_{\rm eff} &=& -\sum_{m=1}^\infty\frac{2T^4}{\pi^2m^4} {\rm
tr}\,\{\,\cos(mgA_0/T) - 1\}
\nonumber \\
&=& \frac{T^2}{6}{\rm tr}\,\{(gA_0)^2(1-\frac{g|A_0|}{2\pi T})^2\}\;,
\label{15}
\end{eqnarray}
where $|A_0|$ means to take the absolute value of each (real,
diagonal) element in $A_0$, and the last line of (\ref{15}) is only
true for fields with $g|A_0|\le 2\pi T$.  To find $V_{\rm
eff}^\prime$, we must subtract the contribution of static ($n=0$) one
loop fluctuations from $V_{\rm eff}$.  Although the full effective
potential requires no new renormalization at finite temperature,
singling out the static (or non-static) modes does require an
additional $A_0$ mass counterterm
\cite{nadk}.  We again choose to use zeta function regularization
and find:
\begin{eqnarray}
V_{\rm eff} - V^\prime_{\rm eff} &=& -\frac{T}{(4\pi)^{3/2}}{\rm tr}\,
\frac{d}{d\epsilon}\{\frac{\Lambda^{2\epsilon}}{\Gamma(\epsilon)}
\int dt\,t^{\epsilon -5/2}[\exp(-g^2A_0^2t)-1]\}
_{\epsilon\rightarrow 0}
\nonumber \\
&=& -\frac{g^3 T}{6\pi}{\rm tr}\, |A_0|^3\;,
\label{16}
\end{eqnarray}
where we assumed $\epsilon >3/2$ to do the integral over $t$ and then
continued to $\epsilon=0$ by implicitly subtracting a (temperature
dependent) mass counterterm.  (This technique is analogous to
continuing from dimension $d<2$ to $d=4$ in dimensional mass
regularization.)  Finally, then, for backgrounds with $g|A_0|\le 2\pi
T$, we get:
\begin{equation}
V^\prime_{\rm eff} = \frac{g^2T^2}{6}{\rm tr}\,\{A_0^2
+\frac{g^2}{(2\pi T)^2}A_0^4\}
\label{17}
\end{equation}
This result has been found previously by many authors \cite{land,nadk,rei}.

\section{Conclusion}

Putting everything together and including the tree-level action, we
find the following expression for the effective action of the static
modes of pure gauge SU(N):
\begin{eqnarray}
S_{\rm eff} &=& -\frac{1}{T}\,{\rm tr}\int d^3x
\{\frac{1}{4N}F_{\mu\nu}^2 +{\textstyle\frac{1}{6}\,} g^2T^2A_0^2 +
\frac{g^4}{24\pi^2}A_0^4
\nonumber \\
&-&\frac{g^2\zeta(3)}{64\pi^4T^2}[
{\textstyle\frac{1}{90}\,} ig F_{\mu\nu}F_{\nu\rho}F_{\rho\mu}
-{\textstyle\frac{19}{60}\,}(D_\mu F_{\mu\nu})^2
-{\textstyle\frac{19}{15}\,} ig F_{0\mu}F_{\mu\nu}F_{\nu 0}
\nonumber \\ &\,&\;\;\;\;\;\;
+{\textstyle\frac{1}{30}\,}(D_\mu F_{\mu 0})^2
-{\textstyle\frac{6}{5}\,}igA_0(D_\mu F_{\mu\nu})F_{0\nu}
+{\textstyle\frac{11}{6}\,}g^2A_0^2F_{\mu\nu}^2]\}
\label{18}
\end{eqnarray}
where $g$ is the running coupling, and the trace is over adjoint
representation matrices.  Although there are no time derivatives in
the above equation, for convenience we use the notation $D_0 = -igA_0$
and $(F_{i0})^{ab}=-if^{abc}(D_i^{cd}A_0^d)$.  The top line of this
effective action is well known in the literature
\cite{land,nadk,rei,kark}, but the second and third lines contain new
terms which may lead to improved agreement of the effective theory
with full 4-dimensional QCD at temperatures closer to the phase
transition.
\\[4ex]
{\noindent\large\bf Acknowledgements}

I wish to thank U. Heinz and M. Oleszczuk as well as others in the
Regensburg Nuclear Theory group for interesting and fruitful
discussions.  Financially, this work was supported by BMFT.

\newpage
\setcounter{section}{1}
\renewcommand{\theequation}{\Alph{equation}}
\setcounter{equation}{0}
{\noindent\Large\bf Appendix}

To get eqn. (\ref{5}), we expand the exponential operator of the ghost
determinant in powers of covariant derivatives, keeping terms
involving up to six covariant derivatives and all terms involving only
$D_0$.  After integrating over $d^3p$, we are left with a large number
of terms which can be regrouped as follows:
\begin{eqnarray}
&\;&180(4\pi t)^{3/2}\int\frac{d^3p}{(2\pi)^3}\,\sum_n\,e^{-p^2t}
\exp[(D^2 +2iD_\alpha p_\alpha)t]
\nonumber \\
&\;&\;\;\;\;
=\sum_n\,e^{-p_0^2t}\{180\exp[(D_0^2+2ip_0D_0)t]
\nonumber \\
&\;&\;\;\;\;
+15t^2[D_\mu,D_\nu]^2
+30t^2(1-2p_0^2t)
([D_i,D_0]^2 +\{[D_i,[D_i,D_0]],D_0\})
\nonumber \\
&\;&\;\;\;\;
+t^3(-6{\cal O}_1 +{\cal O}_2 +4{\cal O}_3
+3{\cal O}_4 + 3{\cal O}_5)
\\
&\;&\;\;\;\;
+t^3(1-2p_0^2t)
(-18{\cal P}_1 -6{\cal P}_2 -6{\cal P}_3
+2{\cal P}_4 +2{\cal P}_5 + 5{\cal P}_6 + 2{\cal P}_7 + 4{\cal P}_8
\nonumber \\
&\;&\;\;\;\;\;\;\;\;\;\;\;\;\;
+{\textstyle\frac{3}{2}\,}{\cal P}_9
+{\textstyle\frac{3}{2}\,}{\cal P}_{10}
+6{\cal P}_{11} +6{\cal P}_{12} -6{\cal P}_{13}
-6{\cal Q}_1 +6{\cal Q}_2
+{\textstyle\frac{3}{2}\,}{\cal Q}_3
-{\textstyle\frac{3}{2}\,}{\cal Q}_4
\nonumber \\
&\;&\;\;\;\;\;\;\;\;\;\;\;\;\;
+{\textstyle\frac{9}{2}\,}{\cal Q}_5
-{\textstyle\frac{9}{2}\,}{\cal Q}_6
+15{\cal Q}_7+3{\cal Q}_8
-12{\cal Q}_9 +12{\cal Q}_{10}+18{\cal Q}_{11}-6{\cal Q}_{12}
+3{\cal Q}_{14})
\nonumber \\
&\;&\;\;\;\;
+2t^3(3-12p_0^2t+4p_0^4t^2)
(-{\cal P}_{13}-3{\cal Q}_9 +3{\cal Q}_{10} +6{\cal Q}_{11}
+3{\cal Q}_{12}+5{\cal Q}_{13} +2{\cal Q}_{14})\}
\nonumber
\label{a1}
\end{eqnarray}
where curly brackets indicate anticommutators, $i$ and $j$ are used for
spatial indices only, and the operators which arise in the sixth order
of the expansion are given below.

The ${\cal O}$ operators were originally presented in in the context
of a zero temperature covariant derivative expansion \cite{dyak}.
These completely gauge invariant and relativistically covariant
commutators are given by:
\begin{eqnarray}
&\,&\hspace*{1.2in}{\cal O}_1 =
[D_\mu,D_\nu][D_\nu,D_\rho][D_\rho,D_\mu]\;\;\;\;\;\;\;
\nonumber \\
&\,&{\cal O}_2 = [D_\mu,[D_\mu,D_\rho]][D_\nu,[D_\nu,D_\rho]]
\;\;\;\;\;\;
{\cal O}_3 = [D_\mu,[D_\nu,D_\rho]][D_\mu,[D_\nu,D_\rho]]
\nonumber \\
&\,&{\cal O}_4 = [D_\mu,[D_\mu,[D_\nu,D_\rho]]][D_\nu,D_\rho]
\;\;\;\;\;\;
{\cal O}_5 = [D_\nu,D_\rho][D_\mu,[D_\mu,[D_\nu,D_\rho]]]
\nonumber
\label{a2}
\end{eqnarray}
Although the ${\cal P}$ operators are not relativistically covariant,
they are completely gauge invariant:
\begin{eqnarray}
&\;& {\cal P}_1 = [D_0,D_i][D_i,D_j][D_j,D_0]\;\;\;\;\;\;\;
{\cal P}_2 = [D_i,D_0][D_0,D_j][D_j,D_i]
\nonumber \\
&\;& {\cal P}_3 = [D_i,D_j][D_j,D_0][D_0,D_i]\;\;\;\;\;\;\;
{\cal P}_4 = [D_0,[D_0,D_i]][D_j,[D_j,D_i]]
\nonumber \\
&\;& {\cal P}_5 = [D_j,[D_j,D_i]][D_0,[D_0,D_i]]\;\;\;\;\;\;\;
{\cal P}_6 = [D_i,[D_i,D_0]][D_j,[D_j,D_0]]
\nonumber \\
&\;&{\cal P}_7 = [D_0,[D_i,D_j]][D_0,[D_i,D_j]]\;\;\;\;\;\;\;
{\cal P}_8 = [D_i,[D_j,D_0]][D_i,[D_j,D_0]]
\nonumber \\
&\;&{\cal P}_9 = [D_0,[D_0,[D_i,D_j]]][D_i,D_j]\;\;\;\;\;\;\;
{\cal P}_{10} = [D_i,D_j][D_0,[D_0,[D_i,D_j]]]
\nonumber \\
&\;&{\cal P}_{11} = [D_i,[D_i,[D_j,D_0]]][D_j,D_0]\;\;\;\;\;\;\;
{\cal P}_{12} = [D_j,D_0][D_i,[D_i,[D_j,D_0]]]
\nonumber \\
&\;&\hspace*{1.2in}{\cal P}_{13} = [D_0,[D_0,D_i]][D_0,[D_0,D_i]]
\nonumber
\label{a3a}
\end{eqnarray}
The ${\cal Q}$ operators are only invariant under spatially dependent
gauge transformations:
\begin{eqnarray}
&\;&{\cal Q}_1 = D_0[D_i,[D_i,D_j]][D_0,D_j]\;\;\;\;\;\;\;
{\cal Q}_2 = [D_0,D_j][D_i,[D_i,D_j]]D_0
\nonumber \\
&\;&{\cal Q}_3 = D_0[D_0,[D_i,D_j]][D_i,D_j]\;\;\;\;\;\;\;
{\cal Q}_4 = [D_i,D_j][D_0,[D_i,D_j]]D_0
\nonumber \\
&\;&{\cal Q}_5 = D_0[D_i,D_j][D_0,[D_i,D_j]]\;\;\;\;\;\;\;
{\cal Q}_6 = [D_0,[D_i,D_j]][D_i,D_j]D_0
\nonumber \\
&\;&{\cal Q}_7 = D_0[D_i,D_j]^2D_0
\hspace*{1in}{\cal Q}_8 = \{D_0,[D_i,[D_i,[D_j,[D_j,D_0]]]]\}
\nonumber \\
&\;&{\cal Q}_9 = D_0[D_0,[D_0,D_i]][D_0,D_i]\;\;\;\;\;\;\;
{\cal Q}_{10} = [D_0,D_i][D_0,[D_0,D_i]]D_0
\nonumber \\
&\;&{\cal Q}_{11} = \{D_0^2,[D_0,D_i]^2\}
\hspace*{1in}{\cal Q}_{12} = D_0[D_0,D_i]^2D_0
\nonumber \\
&\;&{\cal Q}_{13} = D_0\{D_0,[D_i,[D_i,D_0]]\}D_0\;\;\;\;\;\;\;
{\cal Q}_{14} = \{D_0[D_0,[D_0,[D_i,[D_i,D_0]]]]\}
\nonumber
\label{a4}
\end{eqnarray}

When traced and integrated over spacetime, all of the ${\cal O}$
operators can be expressed in terms of ${\cal O}_1$ and ${\cal O}_2$:
\begin{eqnarray}
{\rm tr}\int d^4x\, {\cal O}_3 = -{\rm tr}\int d^4x\, {\cal O}_4
=-{\rm tr}\int d^4x\, {\cal O}_5 = {\rm tr}\int d^4x\,(-4{\cal O}_1
+2{\cal O}_2)
\nonumber
\label{a5}
\end{eqnarray}
Similarly, we can express all of the ${\cal P}$'s in terms of ${\cal
P}_1$, ${\cal P}_4$, ${\cal P}_6$, and ${\cal P}_{13}$:
\begin{eqnarray}
&\,&{\rm tr}\int d^4x\, {\cal P}_2 = {\rm tr}\int d^4x\, {\cal P}_3
={\rm tr}\int d^4x\, {\cal P}_1
\nonumber \\
&\,&{\rm tr}\int d^4x\, {\cal P}_5 = {\rm tr}\int d^4x\, {\cal P}_4
\nonumber \\
&\,&{\rm tr}\int d^4x\, {\cal P}_7 = -{\rm tr}\int d^4x\, {\cal P}_9
=-{\rm tr}\int d^4x\, {\cal P}_{10} = {\rm tr}\int d^4x\, (-4{\cal P}_1
+2{\cal P}_4)
\nonumber \\
&\,&{\rm tr}\int d^4x\, {\cal P}_8 = -{\rm tr}\int d^4x\, {\cal P}_{11}
=-{\rm tr}\int d^4x\, {\cal P}_{12} = {\rm tr}\int d^4x\, (-4{\cal P}_1
+ {\cal P}_4 + {\cal P}_6)\;,
\nonumber
\label{a9}
\end{eqnarray}
and all of the ${\cal Q}$'s in terms of ${\cal Q}_1$, ${\cal Q}_3$,
${\cal Q}_7$, ${\cal Q}_8$, ${\cal Q}_9$, ${\cal Q}_{12}$, and ${\cal
Q}_{13}$, and ${\cal Q}_{14}$:
\begin{eqnarray}
&\,&{\rm tr}\int d^4x {\cal Q}_2\,{\bf 1} =
-{\rm tr}\int d^4x {\cal Q}_1\,{\bf 1}
\nonumber \\
&\,&{\rm tr}\int d^4x {\cal Q}_4\,{\bf 1}
= {\rm tr}\int d^4x {\cal Q}_5\,{\bf 1}
=-{\rm tr}\int d^4x {\cal Q}_6\,{\bf 1}
= -{\rm tr}\int d^4x {\cal Q}_3\,{\bf 1}
\nonumber \\
&\,&{\rm tr}\int d^4x {\cal Q}_{10}\,{\bf 1}
= -{\rm tr}\int d^4x {\cal Q}_9\,{\bf 1}
\;\;\;\;\;\;\;
{\rm tr}\int d^4x {\cal Q}_{11}\,{\bf 1}
= 2{\rm tr}\int d^4x {\cal Q}_{12}\,{\bf 1}\;.
\nonumber
\label{a10}
\end{eqnarray}
When these substitutions are made in (A), the result is eqn. (\ref{5})

\end{document}